\documentclass[]{mem}
\usepackage{natbib}
\usepackage{txfonts}
\usepackage{balance}
\usepackage{graphicx}
\idline{78}{271}

\begin{document}
\def\teff{$T\rm_{eff }$}
\def\kms{$\mathrm {km s}^{-1}$}

\title{Magnetic flux emergence in fast rotating stars}
\author{V.\,Holzwarth }

\institute{Max-Planck-Institut f\"ur Sonnensystemforschung,
Max-Planck-Str.\ 2, 37191 Katlenburg-Lindau, Germany;
\email{holzwarth@mps.mpg.de} }
\offprints{V.\,Holzwarth}

\authorrunning{V.\ Holzwarth}
\titlerunning{Magnetic flux emergence in fast rotating stars}

\abstract{ 
Fast rotating cool stars are characterised by high magnetic activity 
levels and frequently show dark spots up to polar latitudes.
Their distinctive surface distributions of magnetic flux are
investigated in the context of the solar-stellar connection by applying
the solar flux eruption and surface flux transport models to stars with
different rotation rates, mass, and evolutionary stage.
The rise of magnetic flux tubes through the convection zone is primarily 
buoyancy-driven, though their evolution can be strongly affected by the
Coriolis force.
The poleward deflection of the tube's trajectory increases with the
stellar rotation rate, which provides an explanation for magnetic flux
eruption at high latitudes.
The formation of proper polar spots likely requires the assistance of
meridional flows both before and after the eruption of magnetic flux on
the stellar surface.
Since small radiative cores support the eruption of flux tubes at high
latitudes, low-mass pre-main sequence stars are predicted to show high
mean latitudes of flux emergence.
In addition to flux eruption at high latitudes, main sequence
components of close binary systems show spot distributions which are
non-uniform in longitude.
Yet these `preferred longitudes' of flux eruption are expected to vanish
beyond a certain post-main sequence evolutionary stage.
\keywords{stars: magnetic activity -- stars: rotation -- stars:
pre-main sequence -- binaries: close -- magnetohydrodynamics (MHD)}
}

\maketitle{}

\section{Introduction}
The magnetic activity of cool stars has a crucial impact on their
rotational evolution and on their appearance in different wavelength
ranges. 
Detailed observations of solar activity phenomena lead to models for
the cyclic re-generation of magnetic fields through self-sustained
dynamo processes inside the convection zone, which are based on the
interaction of convective motions and (differential) rotation
\citep[][and references therein]{2003A&ARv..11..287O,
2005AN....326..194S}.
Yet a comprehensive understanding of the sub-surface origin of magnetic
flux is only possible when the heterogeneity of other cool stars is
taken into account as well, since the large ranges of stellar rotation
rates, stellar masses, and evolutionary stages provide the testbed
required to verify (or falsify) current theories.

Photometric and spectro-polarimetric observations allow for the
reconstruction of stellar surface brightness and surface magnetic field
distributions \citep[e.g.][and references therein]{2001astr.conf..183C,
2004AN....325..216H}.
The surface maps frequently show distinctive differences compared to
the solar case like huge spots at high latitudes.  In the context of
the solar-stellar connection, these characteristic phenomena are
investigated in the framework of solar flux transport models.
The present review focuses on the storage, transport, and eruption of
magnetic flux in the convective envelope of fast rotating cool stars.

\section{Magnetic flux of fast rotating stars}
Zeeman-broadening of magnetic sensitive lines provides a measure for
the surface-averaged magnetic flux (flux density times filling factor)
of cool stars, which follows roughly a power law, $\Phi\propto
\Omega^{n_\Phi}$, with a rate of increase $ n_\Phi\simeq 1.2$
\citep{2001ASPC..223..292S}; $\Omega$ is the stellar rotation rate.
Excluding targets which might be in the regime of saturated dynamo
operation, \citet{2003ApJ...590..493S} suggest a higher value,
$n_\Phi\simeq 2.8$.
In contrast, observations of young open stellar clusters of different
age indicate a spin-down of stellar rotation due to magnetic braking,
which is consistent with a linear increase of the \emph{open} magnetic
flux \citep[e.g.][and references therein]{2005A&A...444..661H}.
The combination of an empirical relationship between unsigned magnetic
flux and coronal X-ray emission \citep{2003ApJ...598.1387P} with
empirical activity-rotation-relations
\citep[e.g.][]{2003A&A...397..147P} implies an intermediate value
$n_\Phi\sim 2$.

In the course of its 11-year activity cycle, the total spot coverage of
the Sun may reach $0.5\%$ of the visible hemisphere.
In contrast, the spot coverage of rapidly rotating stars can be over
two orders of magnitude larger \citep{2004AJ....128.1802O}.
Whereas sunspots appear within an equatorial belt between about $\pm
35^\mathrm{o}$, fast rotating stars frequently show huge spots at high
and polar latitudes as well (e.g. \citealt{2001MNRAS.324..231B,
2004MNRAS.348.1321B, 2002A&A...389..202O, 2004A&A...417.1047K,
2007MNRAS.tmp.1492J}; see also \citealt{2002AN....323..309S}, and
references therein).
Furthermore, spectro-polarimetric Zeeman-Doppler imaging observations
indicate that the magnetic field pattern of rapidly rotating stars is
characterised by a significant mixture of polarities, which is in
contrast to the unipolar field around the solar poles
\citep{2003MNRAS.345.1145D}.

\section{The Solar Paradigm}
\begin{figure}[t!]
\resizebox{\hsize}{!}{
\includegraphics[width=.24\textwidth]{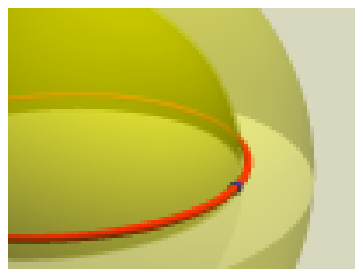}
\includegraphics[width=.24\textwidth]{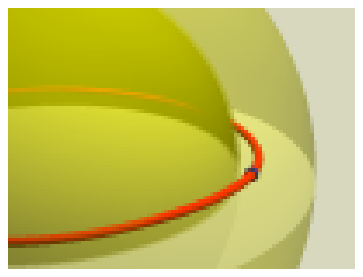}
\includegraphics[width=.24\textwidth]{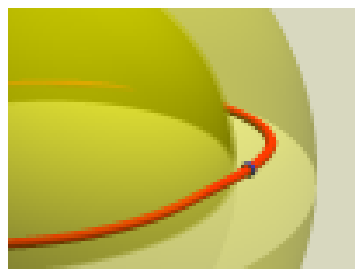}
\includegraphics[width=.24\textwidth]{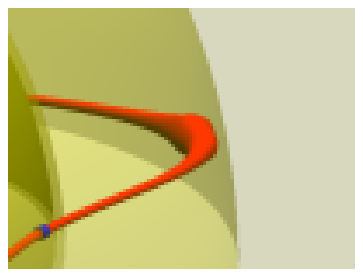}
 }
\resizebox{\hsize}{!}{
\includegraphics[width=.24\textwidth]{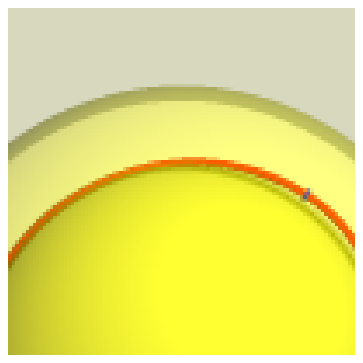}
\includegraphics[width=.24\textwidth]{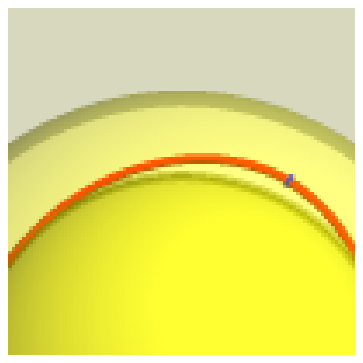}
\includegraphics[width=.24\textwidth]{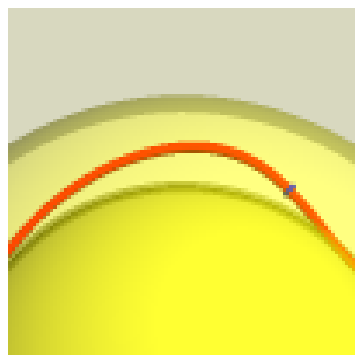}
\includegraphics[width=.24\textwidth]{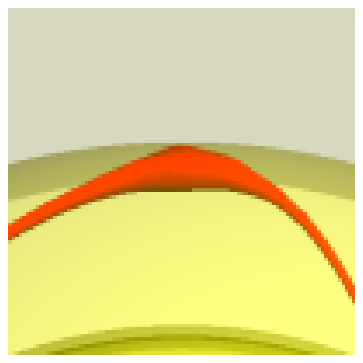} 
}
\caption{\footnotesize Rise of a magnetic flux loop inside the convection 
zone, from the onset of the instability (\emph{left}) to its eruption
at the stellar surface (\emph{right}).
The flux tube radius is shown $5\times$ magnified.
}
\label{solpdm}
\end{figure}

The magnetic field permeating the solar atmosphere is expected to
originate from the bottom of the convection zone.
The field is amplified in the tachocline and stored in the stably
stratified overshoot region at the interface to the radiative core
\citep{1982A&A...113...99V, 1986A&A...166..291M}.
When the field strength is larger than a critical value, perturbations
lead to the formation of rising flux loops (Fig.\ \ref{solpdm}), which
eventually emerge at the surface \citep{1987ApJ...316..788C,
1994ApJ...436..907F, 1994A&A...281L..69S}.

After the dynamical disconnection from its sub-surface roots
\citep{1999SoPh..188..331S, 2005A&A...441..337S}, the surface magnetic
flux feature follows the differential rotation and the meridional flow
in the photosphere.
During its transport toward the pole, surface magnetic flux merges and
annihilates with ambient flux features and eventually dissolves through
diffusion and through the convective turnover of supergranular motions
\citep{DeVore1984, vanBal1998, 2004A&A...426.1075B}.

The decapitated flux tube below the surface disintegrates in
magneto-convective motions and may be transported by meridional
circulations and through convective pumping back to the bottom of the
convection zone for recurrent amplification
\citep{2001ApJ...549.1183T}.

Calculations based on the solar paradigm and carried out in the
framework of the thin flux tube approximation
\citep{1981A&A...102..129S} predict critical field strengths, eruption
latitudes, tilt angles, and proper motions of spots which are in
agreement with observed properties of emerging bipolar spot groups on
the Sun \citep{1993A&A...272..621D, 1994ApJ...436..907F,
1995ApJ...441..886C, 1996A&A...314..503S}.

\section{Flux tubes in fast rotating stars}
Theoretical investigations of magnetic flux eruption in cool stars are
based on analyses of the equilibrium, stability, and rise of flux tubes
for different stellar rotation rates, masses, and evolutionary stages.

\begin{figure}[t!]
\resizebox{\hsize}{!}{\includegraphics[clip=true]{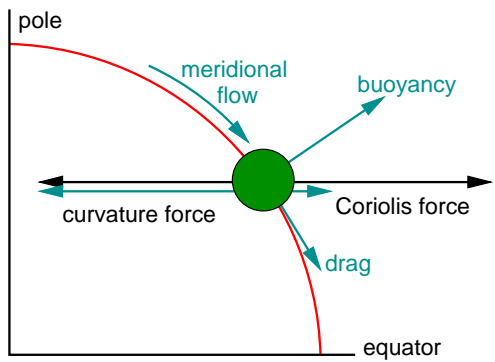}}
\caption{\footnotesize Force balance of a magnetic flux ring in mechanical 
equilibrium in the absence (\emph{black arrows}) and in the presence
(\emph{gray arrows}) of an equatorward meridional flow.}
\label{forces}
\end{figure}

\paragraph{Equilibrium properties}
The magnetic flux tubes are assumed to be initially situated inside the
overshoot region, stored in mechanical equilibrium parallel to the
equatorial plane \citep{1982A&A...106...58S, 1992A&A...264..686M}.
The flux ring is in pressure equilibrium with its environment and, in
the absence of meridional circulations, non-buoyant.
The magnetic tension force pointing toward the axis of rotation is
balanced by the Coriolis force (Fig.~\ref{forces}), which is caused by
an internal prograde flow along the flux ring.

A relative motion between a flux tube and its environment perpendicular
to the tube's axis gives rise to a hydrodynamic drag.
In the presence of an equatorward meridional flow, the resulting drag
is balanced by assuming a buoyant flux tube with a somewhat lower
internal flow velocity \citep{1988ApJ...333..965V,
2006MNRAS.369.1703H}

\begin{figure}[t!]
\resizebox{\hsize}{!}{\includegraphics[clip=true]{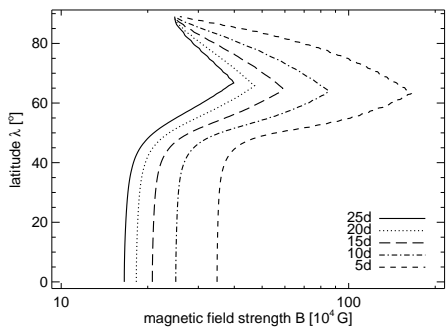}}
\caption{\footnotesize Buoyancy-driven instability with a growth time
of $100\,{\rm d}$ for pre-main sequence stars ($M= 1\,{\rm M_\odot}, R=
1.4\,{\rm R_\odot}, t= 4.7\,{\rm Myr}$) with different rotation periods.}
\label{stabdiag}
\end{figure}

\paragraph{Stability properties}
In the case of the Sun, magnetic flux tubes in the overshoot region are
(linearly) stable against perturbations for field strengths $\lesssim
10^5\,{\rm G}$ \citep{1982A&A...106...58S, 1995GAFD...81..233}.
Beyond that critical value, buoyancy-driven instabilities
\citep{1966ApJ...145..811P} lead to the onset of rising flux loops with
characteristic growth times of less then a few hundred days.
Fast stellar rotation increases the stability of a flux ring, since its
enhanced angular momentum hampers perturbations perpendicular to the
rotation axis.
For otherwise unchanged equilibrium conditions, higher field strengths
are required to obtain buoyancy-driven instabilities with comparable
growth times (Fig.\ \ref{stabdiag}).

\begin{figure}[t!]
\resizebox{\hsize}{!}{
 \includegraphics[clip=true]{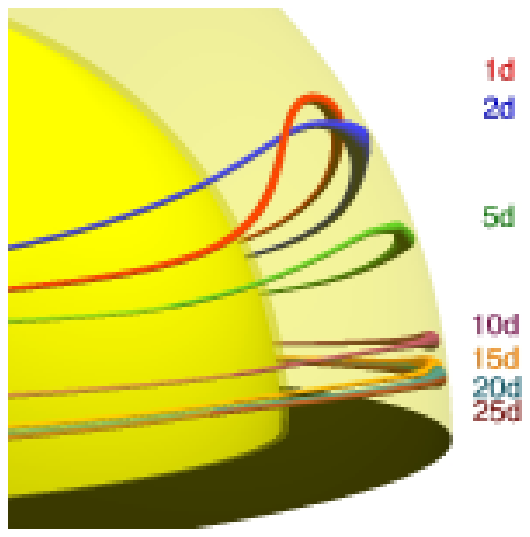} 
 \includegraphics[clip=true]{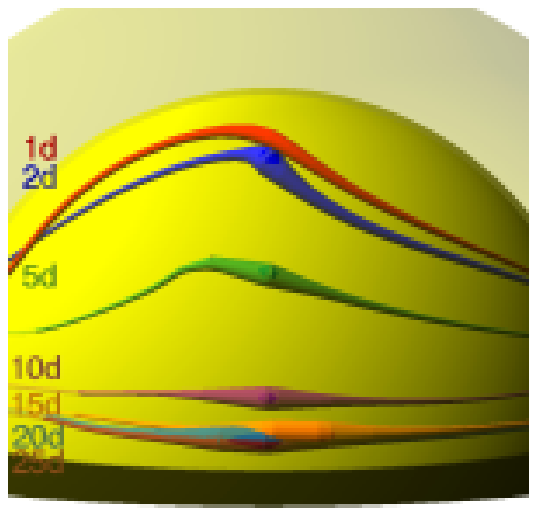}
}
\caption{\footnotesize Poleward deflection (\emph{left}) and tilt angle
(\emph{right}) of erupting flux tubes.
All flux tubes start with the same initial conditions ($B_0= 2\cdot
10^5\,{\rm G}, \lambda_0= 5^\mathrm{o}, r_0= 5.07\cdot 10^{10}\,{\rm
cm}$).}
\label{snap}
\end{figure}

\paragraph{Eruption properties}
The eruption latitude of magnetic flux loops is mainly determined by
the ratio between magnetic buoyancy and Coriolis force.
Magnetic buoyancy is sustained through a net downflow of plasma inside
the flux tube from its crest into the lower segments remaining in the
overshoot region.
The downflow and the density contrast increase with the magnetic field
strength.
If the rise is dominated by magnetic buoyancy, the trajectory will be
radial and the eruption latitude similar to the initial latitude of the
flux ring in the overshoot region.
The azimuthal flow velocity of plasma within a rising flux loop
decreases, owing to the (quasi-)conservation of angular momentum.
The associated decrease of the Coriolis force reduces the outward
directed net force perpendicular to the rotation axis and entails a
deflection of the loop's trajectory to higher latitudes.
The dependence of the Coriolis force on the stellar rotation rate
causes the poleward deflection (of flux tubes with comparable field
strength and equilibrium position) to be larger in more rapidly
rotating stars (Fig.\ \ref{snap}, left).

The deflection mechanism applies to each tube segment as well.
The downflow velocity in the leading leg (relative to the direction of
rotation) is larger than the upflow velocity in the following leg, so
that the former is less deflected toward the pole than the latter.
The asymmetric deflection of the two legs causes a twist of the rising
loop and a tilt of the emerging bipolar spot group at the surface with
respect to the East-West direction (Fig.\ \ref{snap}, right).

\begin{figure}[t!]
\resizebox{\hsize}{!}{\includegraphics[clip=true]{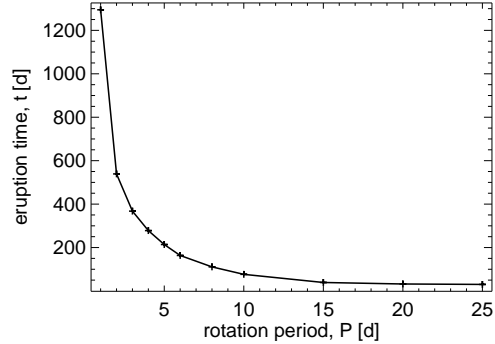}}
\caption{\footnotesize Eruption times of magnetic flux tubes.
The solar-like stellar structure and initial equilibrium conditions
($B_0= 2\cdot 10^5\,{\rm G}, \lambda_0= 5^\mathrm{o}, r= 5.07\cdot
10^{10}\,{\rm cm}$) are the same for all flux tubes.}
\label{erupti}
\end{figure}

Both the poleward deflection and the tilt angle of emerging bipoles
depend on the ratio between buoyancy and Coriolis force, i.e.\ eruption
timescale and rotation period, respectively.
The eruption times of magnetic flux tubes with the same initial field
strength and equilibrium position increase with the rotation rate
(Fig.\ \ref{erupti}), which confirms the influence of the enhanced
angular momentum on the sub-surface evolution of magnetic flux
indicated by the linear stability analysis.

\section{Formation of polar spots}

\begin{figure}[t!]
\resizebox{\hsize}{!}{ 
 \includegraphics[clip=true]{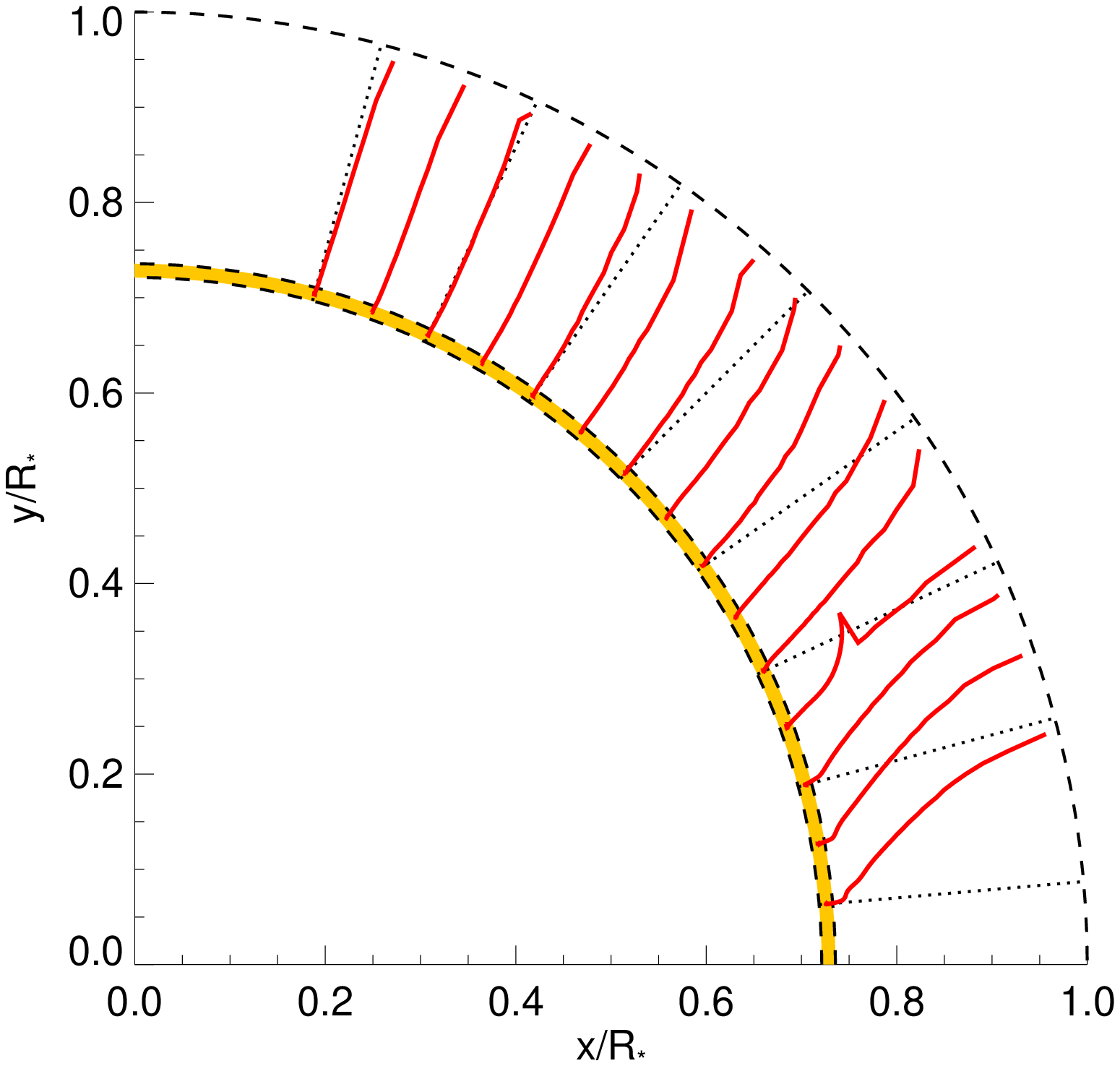}
 \includegraphics[clip=true]{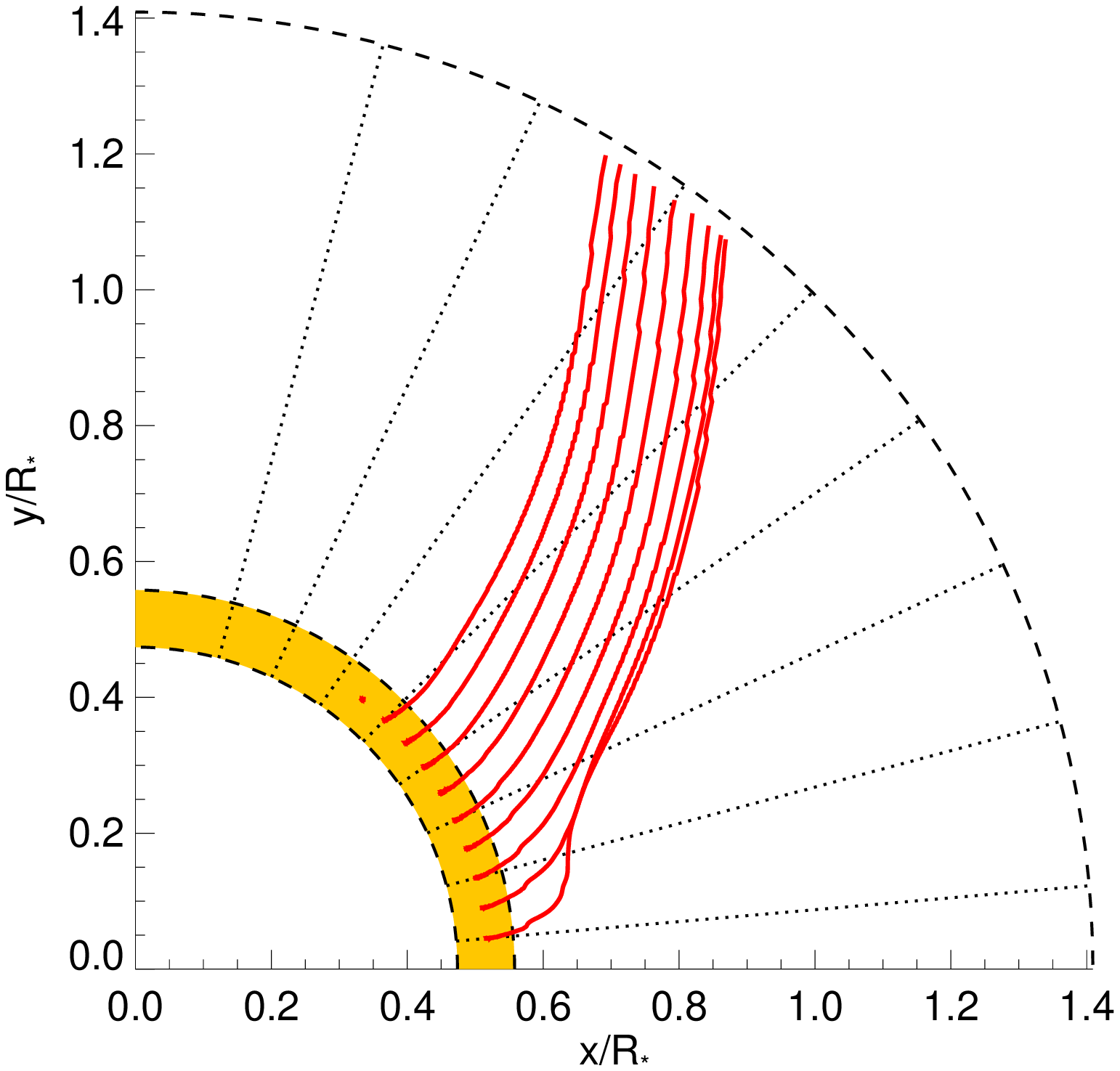} 
}
\caption{\footnotesize Flux loop trajectories in a $1\,{\rm M_\odot}$-main
sequence star (\emph{left}) and a pre-main sequence star (\emph{right})
of age $4.7\,{\rm Myr}$.
The initial field strengths are $B_0= 22\cdot10^4\,{\rm G}$ in the
former case and $B_0= 30\cdot10^4\,{\rm G}$ in the latter; the stellar
rotation period is $P= 6\,{\rm d}$.}
\label{defl}
\end{figure}

\paragraph{Poleward deflection}
In fast rotating stars, the magnetic field strengths required for the
onset of buoyancy-driven instabilities are significantly higher than in
the solar case, which would imply a dominance of magnetic buoyancy and
radial trajectories.
Yet higher initial field strengths entail higher internal flow
velocities to achieve mechanical equilibrium, which in conjunction with
the large rotation rate increase the Coriolis force considerably.
The resulting strong poleward deflection of rising flux loops (Fig.\
\ref{defl}, left) provides an explanation for the occurrence of flux
eruption at high latitudes \citep{1992A&A...264L..13S,
1997ApJ...484..855B}.

\paragraph{Meridional circulation}
Since the poleward deflection decreases for larger initial latitudes,
the formation of polar spots through bona fide flux eruption would
require the presence (and possibly generation) of large amounts of
magnetic flux at very high latitudes in the lower part of the
convection zone.
Although this possibility can a priori not be ruled out, it is more
likely that the formation of polar spots is supported by an additional
poleward transport of magnetic flux through meridional flows.

\begin{figure}[t!]
\resizebox{\hsize}{!}{\includegraphics[clip=true]{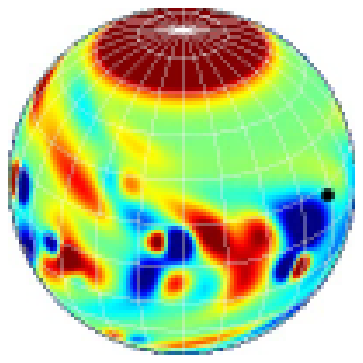}
\includegraphics[clip=true]{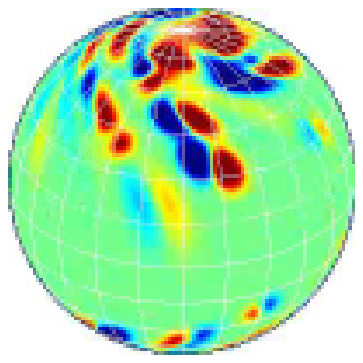} }
\caption{\footnotesize Surface distributions of the radial magnetic
field, assuming solar-like surface transport properties. 
The assumption of a 30 times solar flux eruption rate (\emph{left})
yields a unipolar field at high latitudes, whereas the additional
assumption of a larger latitudinal range of flux eruption and a fast
poleward meridional flow yields a mixture of polarities (\emph{right}).
From \citet{2004MNRAS.354..737M}.}
\label{surflu}
\end{figure}

The attempt to simulate the formation of polar spots on the basis of a
solar surface flux transport model with a $30\times$ larger flux
emergence rate generates unipolar magnetic flux at the poles (Fig.\
\ref{surflu}, left), which disagrees with the mixture of polarities
observed on rapidly rotating stars \citep{2001ApJ...551.1099S,
2003MNRAS.345.1145D}. 
The additional assumption of a larger latitudinal range of flux
emergence and a fast poleward meridional flow yields an intermingling
of polarities (Fig.\ \ref{surflu}, right), which is in qualitative
agreement with observations \citep{2004MNRAS.354..737M}.
The assumed meridional flow velocities ($\gtrsim 100\,{\rm m/s}$) are
significantly high than in the solar case ($11\,{\rm m/s}$).

\begin{figure}[t!]
\resizebox{\hsize}{!}{\includegraphics[clip=true]{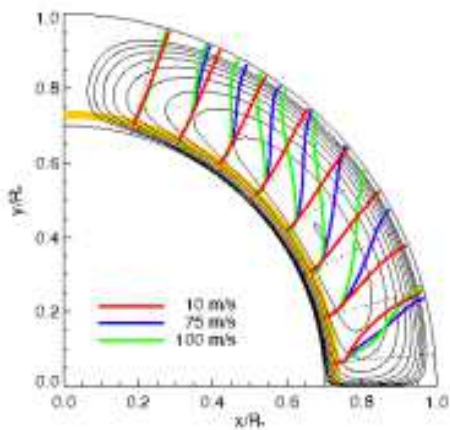} }
\caption{\footnotesize Trajectories of rising flux loops with initial
field strength $15\cdot10^4\,{\rm G}$ and initial tube radius $100\,{\rm
km}$.  The stellar rotation period is $6\,{\rm d}$.
The meridional flow is poleward at the surface and equatorward at the
bottom of the convection zone.
From \citet{2006MNRAS.369.1703H}.}
\label{lattraj}
\end{figure}

\begin{figure}[t!]
\resizebox{\hsize}{!}{ \includegraphics[clip=true]{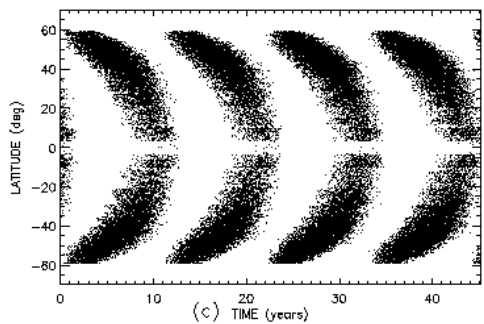} }
\caption{\footnotesize Stellar butterfly diagram in the presence of a
meridional circulation with $100\,{\rm m/s}$ flow velocity.
From \citet{2006MNRAS.369.1703H}.}
\label{butterfly}
\end{figure}

Strong meridional circulations increase the pre-eruptive poleward
deflection of rising flux tubes (Fig.\ \ref{lattraj}).
The deflection is strongest for flux tubes originating from mid
latitudes with low field strengths and small cross sections, which
renders the wings of predicted stellar butterfly diagrams distinctively
convex (Fig.\ \ref{butterfly}).
The enhanced pre-eruptive poleward deflection explains the required
larger range of flux emergence latitudes \citep{2006MNRAS.369.1703H}.

\paragraph{Influence of stellar structure}
The marginal case of poleward deflection corresponds to the rise of a
flux ring parallel to the stellar rotation axis
\citep[e.g.][]{1997ApJ...484..855B}, which enables flux emergence at
high latitudes, depending on the relative size of the radiative core.
This dependence on the stellar structure makes rapidly rotating
pre-main sequence stars optimal candidates for polar spots
\citep{2000A&A...355.1087G, 2002AN....323..395G}.

\begin{figure}[t!]
\resizebox{\hsize}{!}{\includegraphics[clip=true]{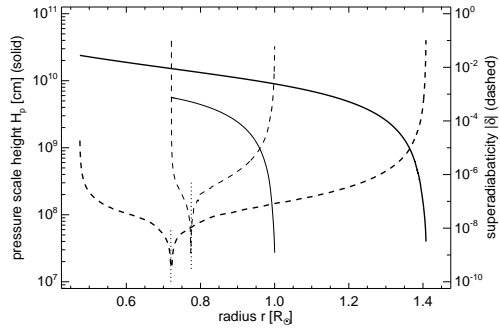}}
\caption{\footnotesize Pressure scale height (\emph{solid}) and
superadiabaticity, $\delta= \nabla - \nabla_\mathrm{ad}$, in the convection
zone of a $4.7\,{\rm Myr}$ old pre-main sequence star (\emph{thick lines})
and a solar-like main sequence star (\emph{thin lines}); to the left of
the vertical \emph{dotted lines} the stratification is subadiabatic
(i.e.\ $\delta<0$).}
\label{stratcmp}
\end{figure}

The poleward deflection is supported by the stratification of the
convection zone of young stars, which is characterised by larger
pressure scales heights and lower superadiabaticities than in a main
sequence star (Fig.\ \ref{stratcmp}).
The smaller magnetic buoyancy enhances the influence of the Coriolis
force, so that the poleward deflection is increased (Fig.\ \ref{defl},
right).
The larger stellar radii and deeper convection zones of pre-main
sequence stars also imply longer rise times, during which the poleward
deflection cumulates to higher eruption latitudes.

\section{Distributions of flux eruption}

\begin{figure}[t!]
\resizebox{\hsize}{!}{ \includegraphics[clip=true]{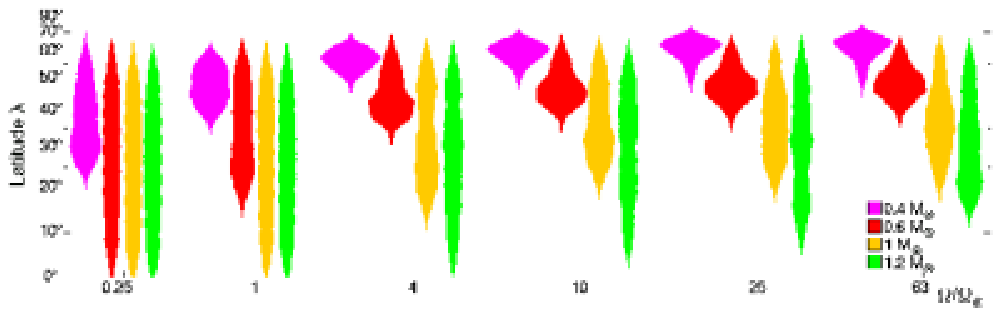} }
\resizebox{\hsize}{!}{ \includegraphics[clip=true]{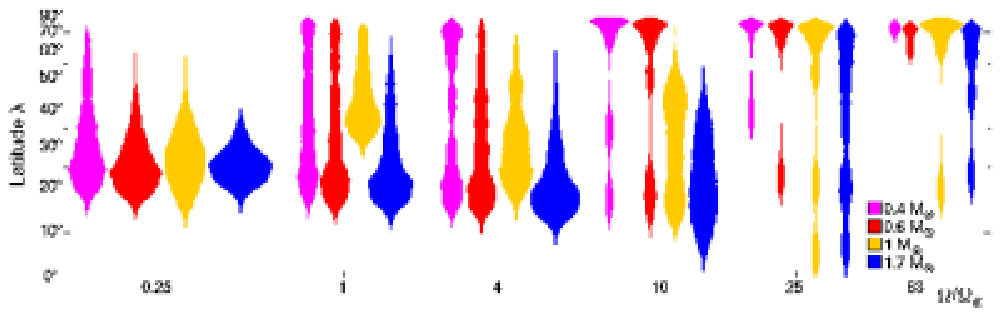} }
\caption{\footnotesize Latitudinal probability distributions of flux
eruption in stars of different stellar mass on the zero-age main
sequence (\emph{top}) and shortly after the Hayashi phase
(\emph{bottom}).
From \citet{dissgranzer00}.}
\label{lpd}
\end{figure}

\paragraph{Young stars}
Given the influence of the stellar structure and stratification on the
formation of polar spots, the mean latitude of magnetic flux eruption
is predicted to increase for stars of lower mass and of earlier
evolutionary stage \citep{2000A&A...355.1087G, 2002AN....323..395G}.
Comparing pre-main sequence stars of similar evolutionary stage,
latitudinal probability distributions based on simulations of erupting
flux tubes show that for lower mass stars the mean flux eruption
latitude is very high already for rotation periods of a week, whereas
for higher mass stars flux emergence can be expected to proceed at low
and intermediate latitudes for rotation rates in the saturated regime
(Fig.\ \ref{lpd}, top).
For even earlier evolutionary stages, magnetic flux emergence can still
occur down to low latitudes through a different eruption mechanism
(Fig.\ \ref{lpd}, bottom).
When the relative size of the radiative core is very small, unstable
magnetic flux rings slip at the bottom of the convection zone to the
pole and detach from the overshoot region.
If the instability is axial symmetric, the flux ring can rise along the
rotation axis and erupt at the pole; if the instability is
non-axisymmetric, the flux tube drifts toward the equator and
eventually emerges at lower latitudes \citep{2000A&A...355.1087G}.
The predicted eruption of magnetic flux down to low latitudes even in
the case of rapid rotation is in agreement with observations.

\begin{figure}[t!]
\resizebox{\hsize}{!}{ \includegraphics[clip=true]{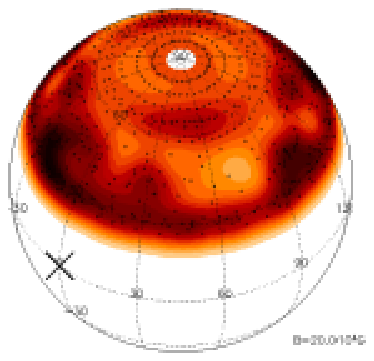}
\includegraphics[clip=true]{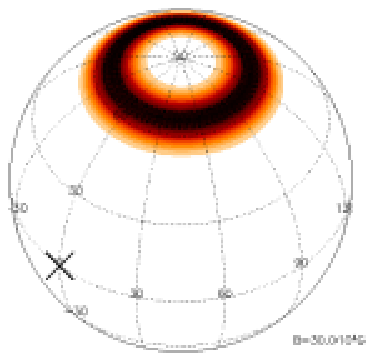} }
\caption{\footnotesize Surface distributions of erupting flux tubes in
a main sequence component (\emph{left}) and a post-main sequence
component (\emph{right}) of a binary system with two $1\,{\rm
M_\odot}$-stars and a $2\,{\rm d}$ rotation period; the cross marks the
direction to the companion star.
From \citet{2004AN....325..408H}.}
\label{bindist}
\end{figure}

\paragraph{Binary stars}
Close binaries with cool stellar components like RS CVn- or BY
Dra-systems show strong magnetic activity, since tidal interactions
maintain high rotation rates against magnetic braking.
The presence of the companion star gives rise to tidal forces and a
deformation of the stellar structure, which in the lowest order of
approximation is $\pi$-periodic in longitude.
The tidal effects modify the equilibrium, stability, and eruption
properties of magnetic flux tubes \citep{2003A&A...405..291H,
2003A&A...405..303H}.
Albeit the tidal perturbations are rather small, their resonant
interaction with double-looped (i.e. roughly $\pi$-periodic) flux tubes
result in considerable non-uniformities in the surface probability
distributions of magnetic flux eruption (Fig.\ \ref{bindist}, left).
The orientation of the resulting $\pi$-periodic `preferred longitudes'
of flux eruption depends on the initial field strength and latitude of
the original flux ring in the overshoot region.

Beyond a certain post-main sequence evolutionary stage the different
stellar structure and stratification change the stability properties of
flux rings.
Double-loop flux tubes are no longer the dominant mode of flux eruption
but superseded by single-loop flux tubes \citep{2004AN....325..408H}.
Since the latter are incongruent with the $\pi$-periodicity of the
tidal interaction, they are less susceptible to the presence of the
companion star.
The resulting surface probability distribution is almost axial
symmetric, showing hardly any signs of preferred longitudes (Fig.\
\ref{bindist}, right).
Owing to the rapid rotation, flux eruption remains to occur at high
latitudes.

\section{Summary and Conclusion}
High-latitude spots on fast rotating stars are ascribed to the combined
pre-eruptive and post-eruptive poleward transport of magnetic flux
originating from the bottom of the convection zone.
The pre-eruptive poleward deflection and tilt of emerging spot groups
is mainly determined through the ratio between magnetic buoyancy and
Coriolis force, which depend on the stellar structure and rotation
rate, respectively.
Since the deflection typically increases with increasing stellar
rotation rate and decreasing size of the radiative core, rapidly
rotating pre-main sequence stars are expected to have high mean
latitudes of flux eruption, though low latitude spots are still
possible.

The identification of characteristic activity features on fast rotating
stars may provide the key for our understanding of essential dynamo
processes.
Solar activity models describing the pre-eruptive and post-eruptive
transport of magnetic flux have demonstrated their applicability in the
investigations of stellar activity signatures and their dependence on
stellar structure and rotation rate.
Yet there are activity phenomena unaccounted for by the current flux
eruption model, for example, regarding the high activity levels of
fully convective low-mass stars \citep[e.g.][]{2006Sci...311..633D} or
the tentative signs for preferred longitudes on some single stars
including the `flip-flop' phenomenon \citep{1994A&A...282L...9J,
2001A&A...379L..30K, 2004SoPh..224..123B}.

\begin{acknowledgements}
The author thanks the organisers for the invitation to present this
paper, and Drs.\ T.\ Granzer and D.\ H.\ Mackay for providing several
images.
\end{acknowledgements}

\bibliographystyle{aa}

\end{document}